\documentclass[conference]{IEEEtran}

\IEEEoverridecommandlockouts

\usepackage{cite}
\usepackage{amsmath,amssymb,amsfonts}
\usepackage{siunitx}
\usepackage{algorithmic}
\usepackage{graphicx}
\usepackage{textcomp}
\usepackage{xcolor}
\usepackage{booktabs}
\usepackage{tabularx}
\usepackage[para]{threeparttable}
\usepackage{multirow}
\usepackage[normalem]{ulem}
\usepackage{float}
\usepackage[english]{babel}
\usepackage{hyperref}
\usepackage{array}
\usepackage{subcaption}

\useunder{\uline}{\ul}{}

\def\BibTeX{{\rm B\kern-.05em{\sc i\kern-.025em b}\kern-.08em
    T\kern-.1667em\lower.7ex\hbox{E}\kern-.125emX}}

\graphicspath{{./images/}}

\addto\extrasenglish{%
}

\newcommand{\ra}[1]{\renewcommand{\arraystretch}{#1}}

\lccode`8=`8
\begin{document}

\title{
ESResNe(X)t-fbsp: Learning Robust Time-Frequency Transformation of Audio
\thanks{This work was supported by the TU Kaiserslautern CS PhD scholarship program and the BMBF project ExplAINN (Grant 01IS19074).}
}

\author{
    \IEEEauthorblockN{
        Andrey Guzhov\textsuperscript{1,2},
        Federico Raue\textsuperscript{1},
        J{\"o}rn Hees\textsuperscript{1},
        Andreas Dengel\textsuperscript{1,2}
    }
    \IEEEauthorblockA{
        \textit{\textsuperscript{1}DFKI GmbH} \\
        \textit{\textsuperscript{2}TU Kaiserslautern} \\
        Kaiserslautern, Germany \\
        \url{firstname.lastname@dfki.de}
    }
}

\maketitle

\begin{abstract}
Environmental Sound Classification (ESC) is a rapidly evolving field that recently demonstrated the advantages of application of visual domain techniques to the audio-related tasks.
Previous studies indicate that the domain-specific modification of cross-domain approaches show a promise in pushing the whole area of ESC forward.

In this paper, we present a new time-frequency transformation layer that is based on complex frequency B-spline (fbsp) wavelets.
Being used with a high-performance audio classification model, the proposed fbsp-layer provides an accuracy improvement over the previously used Short-Time Fourier Transform (STFT) on standard datasets.
We also investigate the influence of different pre-training strategies, including the joint use of two large-scale datasets for weight initialization: ImageNet and AudioSet.
Our proposed model out-performs other approaches by achieving accuracies of 95.20\,\% on the ESC-50 and 89.14\,\% on the UrbanSound8K datasets.

Additionally, we assess the increase of model robustness against additive white Gaussian noise and reduction of an effective sample rate introduced by the proposed layer and demonstrate that the fbsp-layer improves the model's ability to withstand signal perturbations, in comparison to STFT-based training.
For the sake of reproducibility, our code is made available.
\end{abstract}


\section{Introduction}
\label{sec:intro}

Environmental Sound Classification (ESC) is a challenging task that implies a correct differentiation between sound classes that occur in our everyday life (e.g., ``sneezing'', ``airplane'', ``jackhammer'', ``cat'', ``idling engine'', ``brushing teeth'', ``street music'').
Widely used datasets, such as ESC-50 \cite{piczak2015esc} and UrbanSound8K \cite{salamon2014us8k}, provide a reliable basis to compare a variety of approaches on the ESC-task, which allowed to confirm the advantage of using cross-domain techniques \cite{guzhov2020esresnet}.

Previously, the general trend in the ESC-community was to design audio-domain-specific architectures.
In the last years, however, the focus has shifted to the use of common techniques from other domains, such as the visual one.
Both directions are combined usually with either a raw signal or a pre-computed time-frequency transformation, which is more common.
Learning of a time-frequency transformation in an end-to-end fashion is a rare exception that, however, is able to provide an increase of accuracy \cite{sailor2017convrbm}.
Also, the usage of a weight initialization obtained on large-scale datasets is in alignment with the recent tendencies.
Such a weight transfer is performed usually in either cross- or intra-domain manner only.
Thus, the field of ESC lacks studies on the assessment of effects of a two-stage domain adaptation using large-scale audio datasets.
Besides that, a typical accuracy evaluation of models is being accomplished in ``ideal'' conditions.
Measuring of the influence of a perturbed signal on the predictions of best performing models is not quite usual.

In our work, we propose a new time-frequency transformation layer that adjusts its parameters to the data and is based on complex frequency B-spline wavelets (fbsp-wavelets) and contributes to the out-performance of previous models.
Also, we introduce an additional pre-training step using a large-scale dataset of audio, namely AudioSet \cite{gemmeke2017audioset}, and evaluate its effect on the classification accuracy for randomly and ImageNet-initialized models.
Finally, we assess the dependency of prediction accuracy of our best performing models on two types of signal perturbations: additive white Gaussian noise and reduction of the effective sample rate.

The remainder of this paper is organized as follows.
In \autoref{sec:related} we discuss prior methods and approaches to Environmental Sound Classification.
Then, we describe the model that includes our proposed time-frequency transformation layer based on complex frequency B-spline wavelets in \autoref{sec:model}, its training and evaluation in \autoref{sec:exp_setup} and the obtained results in \autoref{sec:results}.
Finally, we summarize our work and highlight follow-up research directions in \autoref{sec:conclusion}.

\section{Related Work}
\label{sec:related}

In this section, we describe previous work done in the field of Environmental Sound Classification (ESC).
We highlight approaches that were used to solve the ESC-task, in particular: application of one- and two-dimensional Convolutional Neural Networks (CNN) and the use of pre-computed and trainable transformations.

\subsection{Raw Waveform and 1D-CNN}
\label{sec:related:1dcnn}
One-dimensional CNNs use a raw audio signal as an input and provide a more natural way to build an audio-domain-related model, in comparison to 2D-CNNs.
Since data pre-processing is not needed in such cases, these models provide an out-of-the box learning of a time-frequency transformation \cite{tokozume2017envnet, tokozume2017envnetv2}.
Further enhancement of one-dimensional CNNs was performed in two directions.
One was the operation on different time scales \cite{zhu2018multires}, while the other implied the use of an input layer initialization using gammatone filter banks \cite{slaney1993gammatone, abdoli2019cnn1d} as a starting point for the training.
In this work, we follow a similar direction in application to a two-dimensional CNN, introducing the learning of a time-frequency transformation that is based on the complex fbsp-wavelet filter bank \cite{teolis1998wavelets}.

\subsection{Time-Frequency Representation and 2D-CNN}
\label{sec:related:2dcnn}
The use of image-domain-related CNNs in conjunction with a pre-computed time-frequency representation is a more common setup to solve audio-related tasks.
For the \mbox{ESC-50} dataset, the baseline was set by a model that is referred to as Piczak-CNN \cite{piczak2015cnn}.
The architecture of the Piczak-CNN followed its custom design and was combined with Mel-scaled power spectrograms \cite{volkmann1937mel} .
Later, the follow-up models were based on the Short-Time Fourier Transform (STFT) \cite{allen1977stft} derived data representations (e.g., \cite{salamon2017cnn}, \cite{arnault2020urban}, \cite{kumar2020weanet}, \cite{palanisamy2020densenet}, \cite{zhang2018mixup}, \cite{zhang2019crnn}) or on sophisticatedly designed filter banks (e.g., \cite{tak2017pefbe}).

However, the use of log-power spectrograms without modifications allowed the ESResNet model \cite{guzhov2020esresnet} to achieve state-of-the-art accuracy using a general-purpose visual CNN, suggesting that the Short-Time Fourier Transform itself provides a good representation that can serve as an initial state for a trainable filter bank.

\subsection{Trainable Filter Bank and 2D-CNN}
\label{sec:related:fb_2dcnn}
Currently, in the field of Environmental Sound Classification, two-dimensional CNNs that involve a trainable time-frequency transformation represent the smallest subset of the models.
This situation was caused mainly by the lack of large-scale audio datasets.
However, a successful model was presented that achieved state-of-the-art performance on the ESC-50 dataset \cite{sailor2017convrbm}.
Since the AudioSet dataset \cite{gemmeke2017audioset} was released, it becomes possible to train powerful audio-domain-related neural networks from scratch, including time-frequency transformations.

Finally, we decided to combine advantages of the pre-computed STFT and the ability to fit such a transformation to the data.
In details, the proposed transformation layer is described in \autoref{sec:model:layer}.

\section{Model}
\label{sec:model}

In this section, we will describe the base ESResNet model \cite{guzhov2020esresnet} and the way how it processes its input, the ResNeXt architecture \cite{xie2017resnext} and the proposed trainable time-frequency transformation based on the complex frequency B-spline wavelets \cite{teolis1998wavelets}.

\subsection{ESResNet}
\label{sec:model:esrn}

\begin{figure}[tbp]
	\centering
	\includegraphics[trim={3cm 1cm 13cm 6.5cm},clip,width=\linewidth]{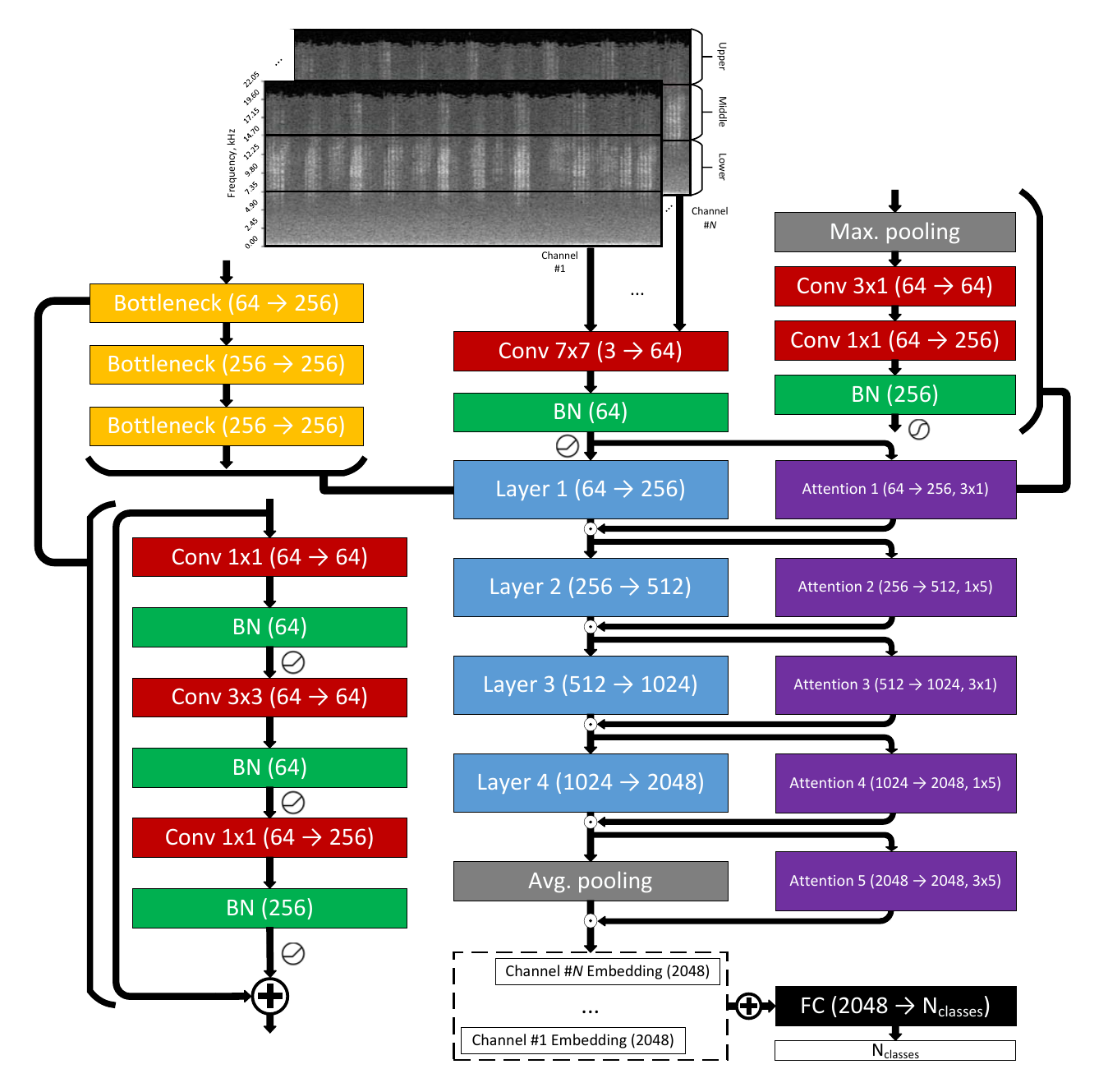}
	\caption{Overview of the ESResNet model. The main branch (central column) consists of the Convolutional layer (red) stacked together with the Batch Normalization layer (green), followed by the residual layers 1\,--\,4 (blue), the Average pooling (gray), and the Fully-Connected layer (black). On the left, the typical structure of a residual layer is presented. Each residual layer consists of the stack of the bottleneck layers (orange) that include \mbox{Conv-BN}-operations applied sequentially and the skip-connection. Rectified Linear Unit (ReLU) serves as an activation function. On the right, structure of the attention blocks is presented. The attention block (violet) is stacked in parallel to the residual layer 1 to 4 or to the average pooling layer. The attention block includes the Max-pooling operation (gray) followed by the depth-wise separable convolution stacked together with batch normalization. The output of the attention block is given by the logistic function.}
	\label{fig:esresnet}
\end{figure}

The ESResNet model was proposed in \cite{guzhov2020esresnet} and combined commonly used visual domain techniques such as a ResNet-based \cite{he2016resnet} backbone, Siamese-like \cite{koch2015siamese} multi-channel processing, and depth-wise separable convolutions \cite{chollet2017xception} together with the computation of log-power spectrograms obtained using Short-Time Fourier Transform and cross-domain transfer learning in application to the ESC-task. The chosen method allowed to achieve state-of-the-art results on the ESC-50 \cite{piczak2015esc} and UrbanSound8K \cite{salamon2014us8k} datasets at the time of publishing.
An overview of the ESResNet's processing pipeline is given by \autoref{fig:esresnet}.

\subsection{ResNeXt}
\label{sec:model:rsnx}
The ResNeXt architecture proposed in \cite{xie2017resnext} is an evolution of the original ResNet model and includes some techniques that were shown successful previously.
In particular, it introduces a so-called ``cardinality" of the residual layers, which refers to a number of paths in the layer.
Given approximately the same number of parameters, the use of the ResNeXt architecture as a backbone for the ESResNet model provides a valuable performance improvement, as described in \autoref{sec:results}.

\subsection{Proposed Layer}
\label{sec:model:layer}
As mentioned in \autoref{sec:model:esrn}, in the ESResNet model \cite{guzhov2020esresnet}, the Short-Time Fourier Transform was applied to transform the input audio to the corresponding time-frequency representation.
This operation can be decomposed into two independent steps: signal framing and a subsequent Discrete Fourier Transform (DFT) of each frame.
In this work, we propose an approach to train a such time-frequency transformation layer that evolves from a sub-optimal state represented by the inverse DFT to an optimal transform in a data-driven manner.

STFT belongs to the family of Fourier-related transforms and is used to determine magnitude and phase of basis sinusoidal frequencies $f_{c}$ at different time points $\tau$ in a time-domain signal $x$ using kernel function $K(n)$.

\begin{equation}
    X(x, \tau) = \sum_{n = -\infty}^{\infty} x[n]w[n - \tau] K_{DFT}^{f_{c}}(n)
    \label{eq:stft}
\end{equation}

\subsubsection{Discrete Fourier Transform}
\label{sec:model:layer:dft}
Discrete Fourier Transform is used mainly to analyze the frequency content of discretized continuous-time signals \cite{oppenheim1999dft}.
The DFT kernel function shown in \autoref{eq:dft_kernel} defines the corresponding time-frequency transformation.

\begin{equation}
    K_{DFT}^{f_{c}}(n) = e^{-2i \pi f_{c} n}
    \label{eq:dft_kernel}
\end{equation}

\subsubsection{Complex Frequency B-Spline Wavelets (fbsp)}
\label{sec:model:layer:fbsp}
The complex frequency B-spline wavelets are compactly defined in the frequency domain and described in terms of order $m$, bandwidth $f_{b}$ and central frequency $f_{c}$ \cite{teolis1998wavelets}.

\begin{equation}
    K_{fbsp}^{m-f_{b}-f_{c}}(n) = \sqrt{f_{b}}\bigg(\frac{f_{b}n}{m}\bigg)^{m}e^{2i \pi f_{c}n}
    \label{eq:fbsp_kernel}
\end{equation}

One can notice that the DFT-kernel defined in \autoref{eq:dft_kernel} represents an inverted in time limiting case of the fbsp-kernel (\autoref{eq:fbsp_kernel}).
Thus, assigning the fbsp-layer's parameters \mbox{$m = 0$} and \mbox{$f_{b} = 1$}, it's initialization becomes identical to the inverse DFT, which makes it possible to start the network's training from a good enough state.

Finally, the transform performed by the proposed fbsp-layer is defined by \autoref{eq:fbsp}.

\begin{equation}
    X(x, \tau) = \sum_{n = -\infty}^{\infty} x[n]w[n - \tau] K_{fbsp}^{m-f_{b}-f_{c}}(n)
    \label{eq:fbsp}
\end{equation}

\subsubsection{fbsp-Specific Loss Term}
\label{sec:model:layer:loss}

In order to regularize weights of the fbsp-filter bank and to preserve the overall signal's energy, we decided to introduce an additional loss term that is specific to the fbsp-layer and, for the N-point fbsp-transform, is described by \autoref{eq:fbsp_loss}.

\begin{equation}
    \mathcal{L}^{fbsp} = \frac{1}{N} \sum_{f_{c} = 0}^{N} (\|K_{fbsp}^{m-f_{b}-f_{c}}\|^2 - 1)^2
    \label{eq:fbsp_loss}
\end{equation}

\section{Experimental Setup}
\label{sec:exp_setup}
In this section, we describe the datasets that were used, training of the models, including our approach to transfer from the visual to the audio domain, and the procedure of robustness' evaluation.

\subsection{Datasets}
\label{sec:exp_setup:datasets}
In this work, we used three audio-domain-related datasets: AudioSet \cite{gemmeke2017audioset}, ESC-50 \cite{piczak2015esc} and UrbanSound8K \cite{salamon2014us8k}.
The AudioSet dataset was used for pre-training of models, whereas the others served for the performance evaluation and assessment of model robustness against additive white Gaussian noise and reduction of an effective sample rate.
Also, the ImageNet \cite{deng2009imagenet} dataset was used as a source of weight initialization in the cross-domain transfer learning.

\subsubsection{AudioSet}
\label{sec:exp_setup:datasets:audioset}
AudioSet was proposed and described in \cite{gemmeke2017audioset}.
It is a large-scale sound dataset that provides $\sim1.8\:\si{\mega}$ sound clips (each $10\;\si{\second}$) organized into 527 classes in a multi-label manner.
The amount of training data allows to use the AudioSet dataset for the initialization of deep neural networks, thus, providing a better initial state for fine-tuning on the \mbox{ESC-50} and UrbanSound8K datasets.

\subsubsection{ESC-50}
\label{sec:exp_setup:datasets:esc50}
The \mbox{ESC-50} dataset consists of 2,000 monaural samples belonging to 50 classes that can be divided into 5 groups, such as \emph{animal} sounds, \emph{natural and water} sounds, \emph{non-speech human} sounds, \emph{interior} and \emph{exterior} sounds \cite{piczak2015esc}.
Samples are distributed equally among classes, thus each category consists of 40 recordings.
Each track has length of $5\;\si{\second}$, the native sample rate is $44.1\:\si{\kilo\hertz}$.
The dataset was divided into 5 folds by its authors that we used in current work to perform our evaluation.

\subsubsection{UrbanSound8K}
\label{sec:exp_setup:datasets:us8k}
The US8K dataset consists of 8,732 samples (both mono and stereo) belonging to 10 classes: ``air conditioner'', ``car horn'', ``children playing'', ``dog bark'', ``drilling'', ``engine idling'', ``gun shot'', ``jackhammer'', ``siren'', and ``street music'' \cite{salamon2014us8k}.
The classes are not balanced in terms of overall recording lengths per class.
Each track has variable length up to $4\;\si{\second}$, the native sample rate varies from $16\:\si{\kilo\hertz}$ to $48\:\si{\kilo\hertz}$.
The dataset was divided into 10 folds by its authors that we used in current work to perform our evaluation.

\subsubsection{ImageNet}
\label{sec:exp_setup:datasets:imagenet}
The ImageNet dataset was proposed and described in \cite{deng2009imagenet}.
It is a large-scale visual dataset that provides more than $1\:\si{\mega}$ images divided into 1000 classes.
As the use of the ImageNet pre-training is beneficial from the performance point of view (e.g., \cite{guzhov2020esresnet,palanisamy2020densenet}), we use ImageNet-trained networks and evaluate the influence of a such initialized models w.r.t. the follow-up training.

\subsection{Hyper-Parameters}
\label{sec:exp_setup:hparams}
In the experiments, we performed training on the AudioSet, ESC-50 and UrbanSound8K datasets from scratch and after the initialization using ImageNet-weights.
The training on the AudioSet dataset was used as an intermediate step, while the two later datasets served as a target for the final performance assessment.

The training on the \mbox{ESC-50} and UrbanSound8K datasets was derived from the one used for the ESResNet model \cite{guzhov2020esresnet}.
According to it, the model was trained for 300 epochs using the Adam optimizer \cite{kingma2014adam} with the learning rate varied from $2.5e-4$ (training from scratch) to $2.5e-5$ (fine-tuning phase), the exponential decay $\gamma = 0.985$ and the $weight$ $decay$ set to $5e-4$.
Other hyper-parameters such as $\beta_{1}$, $\beta_{2}$ and $\epsilon$ were set to the default values.
The fbsp-variant of our model demonstrated a preference to lower learning rate values, which substantiated the decision to reduce it to $1e-5$ and set $\gamma = 0.99$ during the fine-tuning process.

The pre-training stage on the AudioSet dataset is a modification of the original fine-tuning schema, those optimizer is replaced by Stochastic Gradient Descent (SGD) \cite{polyak1992sgd} with Nesterov's momentum \cite{nesterov1983momentum}.
In comparison to the Adam optimizer, it introduces a smaller number of hyper-parameters, which eased finding a well working combination of them.
For the AudioSet pre-training, the $weight$ $decay$ was used the same way as for the fine-tuning phase, as well as the momentum, which was equal to the $\beta_{1}$ parameter of the Adam optimizer.
The huge amount of training samples in the AudioSet dataset allowed us to reduce the number of training epochs to 5.
The choice of the learning rate was determined by its maximum value that allowed to perform the training successfully, and the value varied from $1.6e-3$ (training from scratch) to $4e-4$ (after ImageNet initialization).

For both setups, cross-entropy served as a loss function.

\subsection{Data Augmentation}
\label{sec:exp_setup:augment}
In order to prevent overfitting and improve the prediction accuracy, we utilized the following data augmentation techniques (for both the AudioSet pre-training and the \mbox{ESC-50}\;/\;UrbanSound8K fine-tuning):

\subsubsection{Time Scaling}
\label{sec:exp_setup:augment:scale}
This method can be considered as a combination of time stretching and pitch shift (see \cite{salamon2017cnn}, \cite{tokozume2017envnetv2}).
While the first one changes the duration of an audio file keeping its spectral characteristic untouched, the later one, in opposite to time stretching, allows to manipulate spectral characteristics and preserve the duration of the track.
Both methods rely on computationally expensive operations, which makes it inefficient to apply them in an on-the-fly manner.
The time scaling was chosen to be applied as an augmentation step because of its computational cheapness and effectiveness \cite{guzhov2020esresnet}.
In this work, the scaling factor was distributed uniformly in the range $[-1.5, 1.5]$.

\subsubsection{Time Inversion}
\label{sec:exp_setup:augment:invert}
Time inversion that was applied in \cite{tokozume2017envnetv2} is an effective data augmentation technique that is related to random flip of images during the training on the visual datasets.
Probability of the inversion was set to $0.5$.

\subsubsection{Random Crop}
\label{sec:exp_setup:augment:crop}
For the scaled audio tracks, there was a requirement to align their lengths in order to process the input through the model.
The use of the random cropping instead of the center one allowed to increase the diversity of the data samples even more, thus, acting as an additional augmentation step.
Audio tracks were cropped to the duration of $10\,/\,5\,/\,4\:\si{\second}$ (\mbox{AudioSet}\;/\;\mbox{ESC-50}\;/\;UrbanSound8K, respectively) if they were longer.
Otherwise, no cropping was performed.
During the evaluation phase, the random cropping was replaced by the center one.

\subsubsection{Random Padding}
\label{sec:exp_setup:augment:pad}
The rationale behind random padding is the same as for the random cropping.
Audio tracks were padded to the duration of $10\,/\,5\,/\,4\:\si{\second}$ (\mbox{AudioSet}\;/\;\mbox{ESC-50}\;/\;UrbanSound8K, respectively) if they were shorter.
Otherwise, no padding was performed.
During the evaluation phase, the random padding was replaced by the center one.

\subsection{Cross- and Intra-Domain Transfer Learning}
\label{sec:exp_setup:transfer_learning}
In order to evaluate the role of the AudioSet weight initialization, we performed a series of experiments that included the AudioSet dataset as a pre-training step.
The ESResNet model as well as its fbsp-variant was evaluated on the \mbox{ESC-50} and UrbanSound8K datasets after the training from scratch, ImageNet initialization, AudioSet pre-training from scratch or after the ImageNet weight transfer.
The ESResNeXt-based models were evaluated after the ImageNet weight initialization and the two-stage transfer learning that included the AudioSet intermediate training after the ImageNet initialization.

Additionally, as we noticed that a completely random initialization of all network components could result in mode collapse due to the additional freedom introduced by the early fbsp-filter bank, in such cases we employed a late unfreeze strategy:
We froze the fbsp-layer's parameters for the first three epochs.
This results in the later parts of the network to be trained based on an STFT-like first layer for the first three epochs (similar to the previous \cite{guzhov2020esresnet}), before then updating all parameters in later epochs based on more meaningful gradients.

The transfer of a model to another dataset was done by replacing its last fully-connected layer (the model's linear classifier) by a randomly initialized one, which output shape suited to the task.

\subsection{Evaluation of Robustness}
\label{sec:exp_setup:robustness}
In order to evaluate robustness of trained models to perturbations in the input signal, we conducted experiments that included the addition of noise to the signal and reduction of information in it.

\subsubsection{Robustness Against Additive White Gaussian Noise}
\label{sec:exp_setup:robustness:awgn}
To assess the model's robustness to additive noise, white Gaussian noise at desired Signal-to-Noise Ratios (SNR) was generated and mixed-up to the audio tracks before performing the forward pass through the model.

\subsubsection{Robustness Against Reduction of an Effective Sample Rate}
\label{sec:exp_setup:robustness:sr}

The ability of model to deal with the reduced effective sample rate was tested using low-pass filtering at different cutoff frequencies.

Signal filtering implies attenuation of unwanted frequency components in the signal while preserving amplitude and phase of desired ones without changes.
A low-pass filter passes frequency components that are lower than a chosen cutoff frequency.
The frequency components that lie in a higher band are being suppressed, and, thus, the unwanted part of the input signal is being weakened.
The exact properties of a digital filter depend on its design and include but are not limited to: cutoff frequency (defines filter's passband), passband ripple, slope, width of transition band, stopband, etc.

In this work, we decided to use a 5\textsuperscript{th} order Butterworth low-pass filter, as it provides maximally flat passband response \cite{butterworth1930filter} and a quick roll-off around its cutoff frequency.
The filter was applied before feeding audio samples to the model.

\section{Results}
\label{sec:results}

\subsection{Pre-Training on AudioSet}
\label{sec:results:audioset}

\begin{table}[tbp]
\begin{threeparttable}[t]
\caption{Evaluation Results of the \mbox{ESResNe(X)t} Model on STFT- and fbsp-Spectrograms on the AudioSet dataset.
We can see that fbsp improves over STFT on Audioset.
}
\label{tbl:results_audioset}
\ra{1.2}
\begin{tabularx}{\linewidth}{Xccc}
\toprule
\multicolumn{1}{l}{\multirow{2}{*}{Model}} & \multicolumn{1}{c}{Input} & \multicolumn{1}{c}{ImageNet} & \multicolumn{1}{c}{Mean Average} \\
 & \multicolumn{1}{c}{Type} & \multicolumn{1}{c}{Initialized} & \multicolumn{1}{c}{Precision} \\
\midrule
 \multicolumn{1}{l}{\multirow{5}{*}{ESResNet}} & \multicolumn{1}{c}{\multirow{2}{*}{STFT}} & & 0.1892 \\
 & & \checkmark & 0.2514 \\
\cmidrule{2-4}
  & \multicolumn{1}{c}{\multirow{2}{*}{fbsp}} & & 0.2394 \\
 & & \checkmark & 0.2616 \\
\cmidrule{1-4}
 \multicolumn{1}{l}{\multirow{3}{*}{ESResNeXt}} & \multicolumn{1}{c}{STFT} & \checkmark & 0.2514 \\
\cmidrule{2-4}
 & \multicolumn{1}{c}{fbsp} & \checkmark & 0.2817 \\
\bottomrule
\end{tabularx}
\end{threeparttable}
\end{table}

In \autoref{tbl:results_audioset}, we present mean Average Precision (mAP) obtained by the variants of our proposed model on the evaluation subset of the AudioSet.
The results include scores of the ESResNet model after the training from scratch as well as after the ImageNet initialization for both STFT- and fbsp-based transformations.
Additionally, the effect of the backbone replacement from \mbox{ResNet-50} to \mbox{ResNeXt-50} is evaluated for the two best performing setups, namely STFT- and fbsp-based models in conjunction with the ImageNet weight transfer.
One can observe that the initialization using ImageNet weights is beneficial for the evaluated ESResNet model, as it provides an steady increase of mAP.

The STFT-based model demonstrated a low sensitivity to the replacement of the backbone from \mbox{ResNet-50} to \mbox{ResNeXt-50}, as the corresponding mAP value does not change (0.2514).
At the same time, the fbsp-layer provided a valuable increase of the mAP from $0.2616$ to $0.2871$.

\begin{figure}[tbp]
    \begin{subfigure}[b]{0.495\linewidth}
	    \centering
    	\includegraphics[trim={0.5cm 0cm 0cm 0.7cm},clip,width=\linewidth]{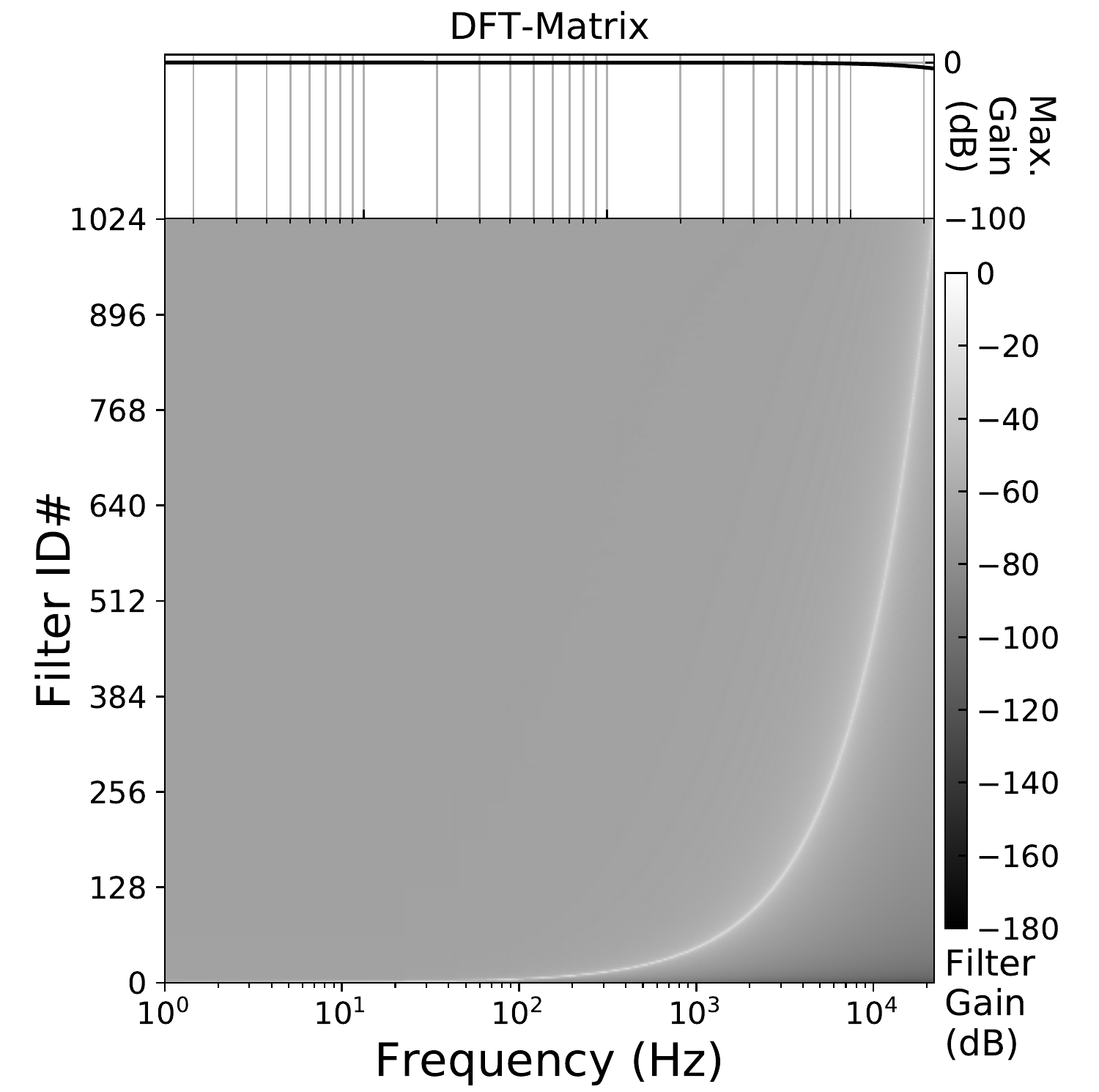}
    	\caption{STFT DFT-Matrix\newline}
    	\label{fig:freq_resp_stft}
    \end{subfigure}
    \hfill
    \begin{subfigure}[b]{0.495\linewidth}
	    \centering
    	\includegraphics[trim={0.5cm 0cm 0cm 0.7cm},clip,width=\linewidth]{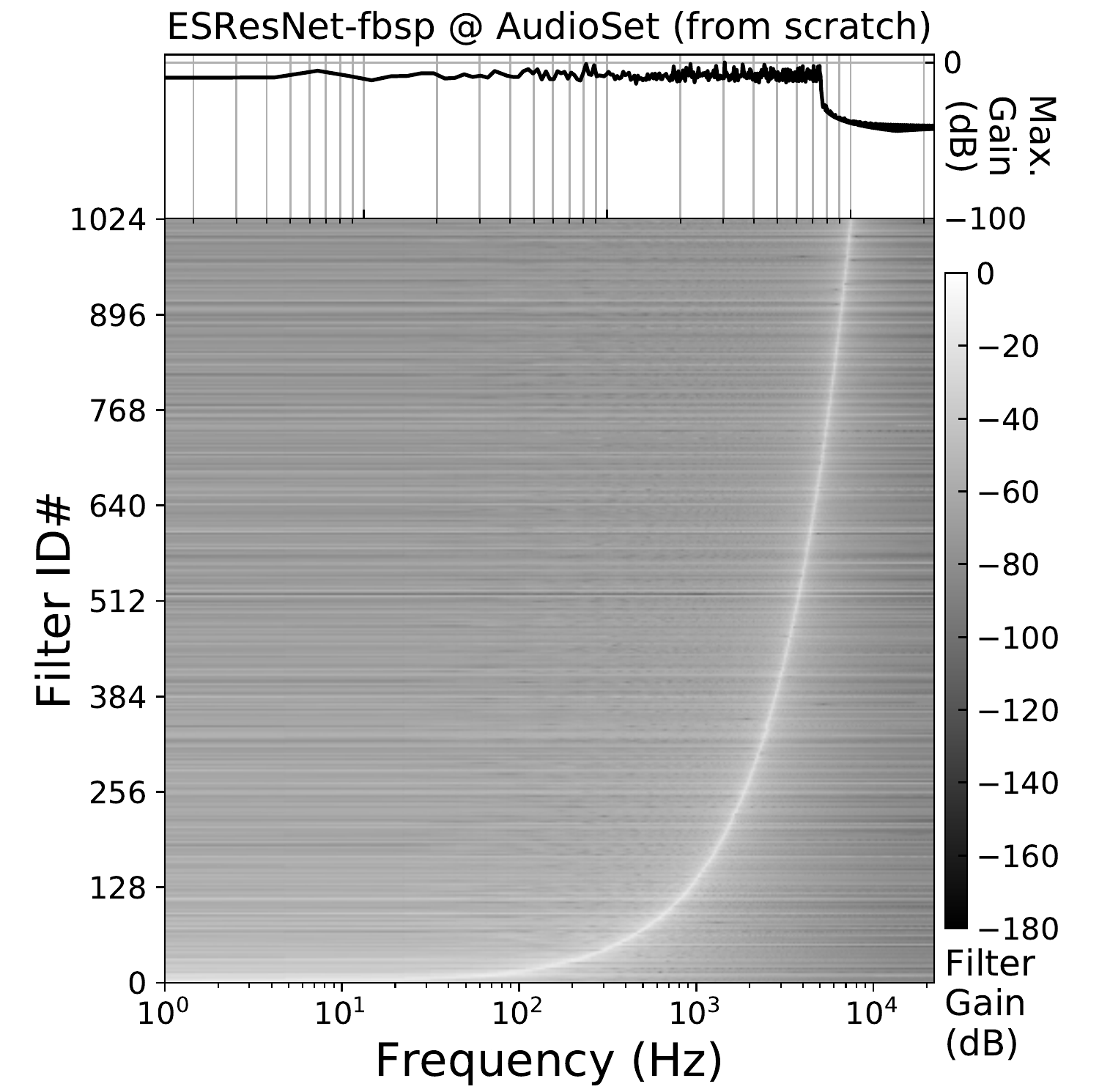}
    	\caption{ESResNet-fbsp @ AudioSet (trained from scratch)}
    	\label{fig:freq_resp_esrn_fbsp_as_fs}
    \end{subfigure}
    \begin{subfigure}[b]{0.495\linewidth}
	    \centering
    	\includegraphics[trim={0.5cm 0cm 0cm 0.7cm},clip,width=\linewidth]{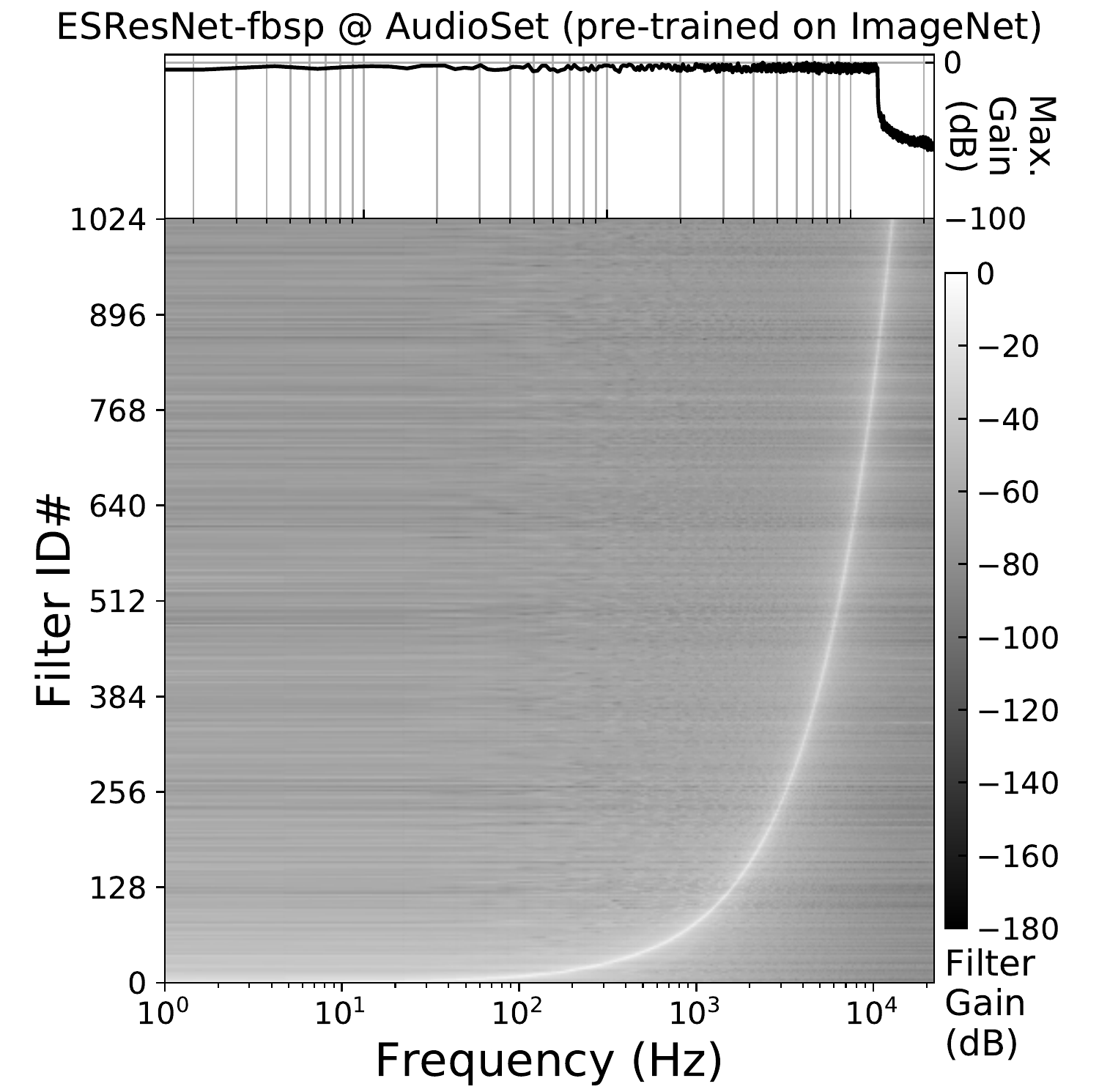}
    	\caption{ESResNet-fbsp @ AudioSet (ImageNet initialization)}
    	\label{fig:freq_resp_esrn_fbsp_as_ptin}
    \end{subfigure}
    \hfill
    \begin{subfigure}[b]{0.495\linewidth}
	    \centering
    	\includegraphics[trim={0.5cm 0cm 0cm 0.7cm},clip,width=\linewidth]{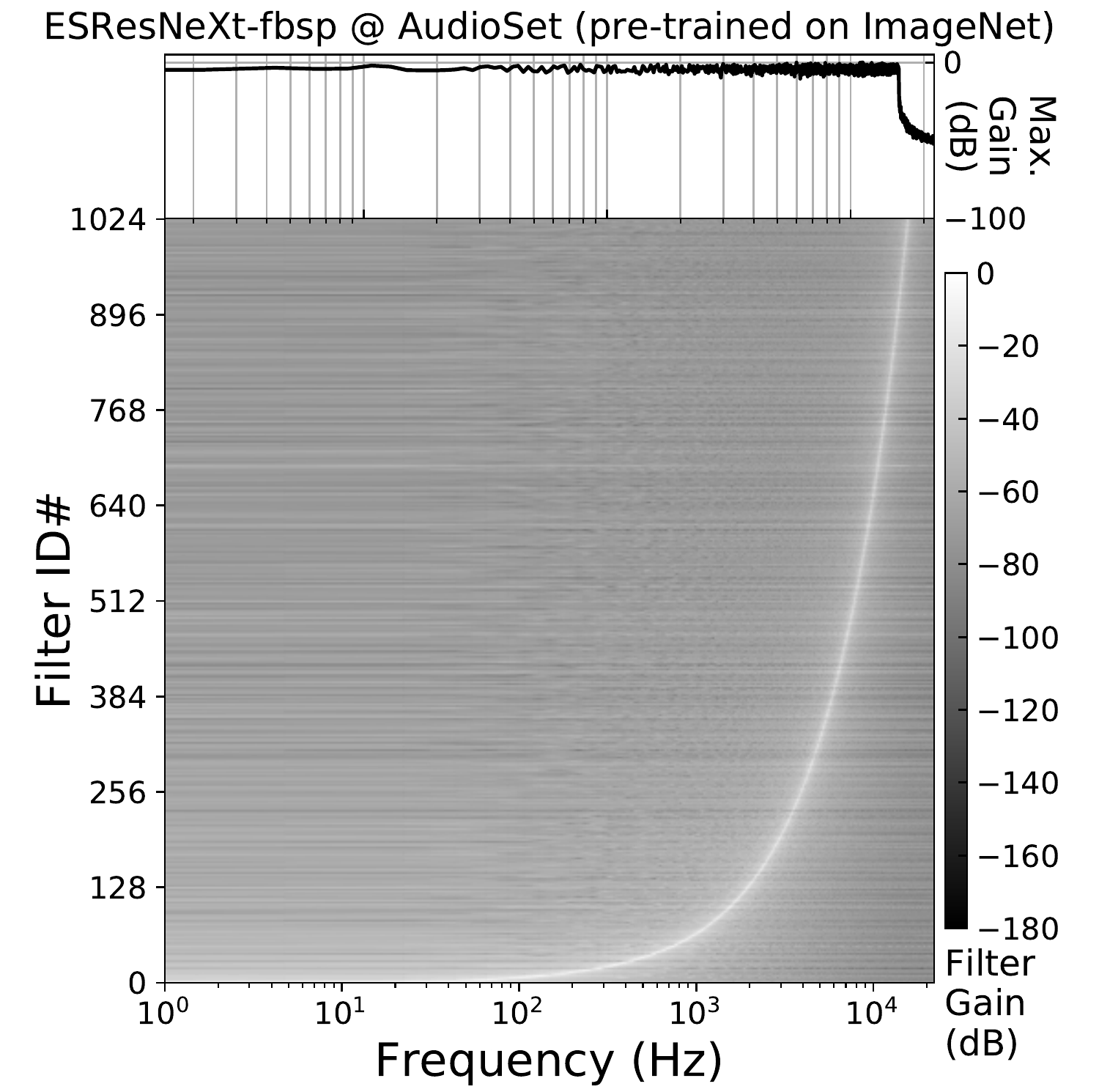}
    	\caption{ESResNeXt-fbsp @ AudioSet (ImageNet initialization)}
    	\label{fig:freq_resp_esrnx_fbsp_as_ptin}
    \end{subfigure}
    
    \caption{Frequency response of filter banks.
    Frequency response shows how the magnitude of each output filter (vertical axis) depends on the frequency of an input signal (horizontal axis).
    A log-scaled plot on top of the corresponding heat map provides an overview of the maximum gain over the frequencies.
    We notice that the fbsp-layers dampen the high frequencies.}
    \label{fig:freq_resp}
\end{figure}

Apart from this, we also computed the frequency responses of the trained on AudioSet fbsp-layers and compared them to the frequency response of an STFT-filter bank.
The frequency response of a filter describes the dependency of the output gain on the frequencies of an input signal.
The DFT-matrix of the STFT-filter bank and its frequency response is shown in \autoref{fig:freq_resp_stft}.
In the top marginal we can observe that the gain is almost flat for the entire frequency band, up to the Nyquist frequency.

In contrast, the frequency responses of the trained fbsp-filter banks (\autoref{fig:freq_resp_esrn_fbsp_as_fs}\:--\:\autoref{fig:freq_resp_esrnx_fbsp_as_ptin}) consist of distinguishable peaks and valleys in the frequency domain.
This fact may indicate that the input signal is redundant w.r.t. sample rate, i.e. the networks learned some sort of signal's decimation.
Also, the trained fbsp-filter banks obtain stopbands in the higher frequency band, which suggests that the filter banks are able to suppress high-frequency components of an input signal's noise.

\subsection{Model Comparison}
\label{sec:results:compare}
In this section, we will discuss the influence of the backbone and the chosen time-frequency transformation on the model's accuracy.

\begin{table}[htbp]
\begin{threeparttable}[t]
\caption{Evaluation Results of the \mbox{ESResNe(X)t} Model on STFT- and fbsp-Spectrograms, accuracy (\%). We can see that
(i) fbsp in general improves over STFT,
(ii) pre-training on both ImageNet and AudioSet improves results, and
(iii) ESResNeXt improves over ESResNet.}
\label{tbl:results_ablation}
\ra{1.2}
\begin{tabularx}{\linewidth}{lXcccc}
\toprule
\multicolumn{1}{l}{\multirow{2}{*}{Model}} & \multicolumn{1}{c}{Input} & \multicolumn{1}{c}{ImageNet} & \multicolumn{1}{c}{AudioSet} & \multicolumn{1}{c}{\multirow{2}{*}{ESC-50}} & \multicolumn{1}{c}{\multirow{2}{*}{US8K}} \\
 & \multicolumn{1}{c}{Type} & \multicolumn{1}{c}{Initialized} & \multicolumn{1}{c}{Pre-Trained} & & \\
\midrule
\multicolumn{1}{r}{\cite{guzhov2020esresnet}} & \multicolumn{1}{c}{\multirow{4}{*}{STFT}} & & & 83.15 & 82.76 \\
\multicolumn{1}{r}{\cite{guzhov2020esresnet}} & & \checkmark & & 91.50 & 85.42 \\
\multicolumn{1}{l}{\multirow{5}{*}{ESResNet}} & & & \checkmark & 92.45 & 87.74 \\
 & & \checkmark & \checkmark & 93.35 & 88.03 \\
\cmidrule{2-6}
 & \multicolumn{1}{c}{\multirow{4}{*}{fbsp}} & & & 86.25 & 83.20 \\
 & & \checkmark & & 91.25 & 85.92 \\
 & & & \checkmark & 92.40 & 88.47 \\
 & & \checkmark & \checkmark & 93.80 & 88.38 \\
\cmidrule{1-6}
\multicolumn{1}{l}{\multirow{5}{*}{ESResNeXt}} & \multicolumn{1}{c}{\multirow{2}{*}{STFT}} & \checkmark & & 91.60 & 86.02 \\
 & & \checkmark & \checkmark & 95.00 & 89.02 \\
\cmidrule{2-6}
 & \multicolumn{1}{c}{\multirow{2}{*}{fbsp}} & \checkmark & & 91.30 & 85.47 \\
 & & \checkmark & \checkmark & \textbf{95.20} & \textbf{89.14} \\
\bottomrule
\end{tabularx}
\end{threeparttable}
\end{table}

\subsubsection{Backbone -- ResNet vs. ResNeXt}
\label{sec:results:compare:backbone}

The influence of the backbone on the model performance was evaluated for two training setups, both of them included the initialization using ImageNet weights (\autoref{tbl:results_ablation}).
For the first one, the models were fine-tuned on the target datasets without intermediate training steps.
The second one included also an AudioSet pre-training.
The use of the ResNeXt instead of the ResNet model as a backbone provided a steady improvement of the prediction accuracy for both setups using STFT for the audio transformation.

Addition of the intermediate AudioSet training also provided a monotonic increase of the performance.
At the same time, when not further pre-training the fbsp-layer with audio data, we can see a minor deterioration of the result for US8K going from ESResNet to ESResNeXt.
However, given such audio-data, we see a strong performance boost of the same setup.

ResNeXt-based variants of the model also demonstrated higher sensitivity to the reduction of the effective sample rate, as will be detailed in \autoref{sec:results:robustness}.

\subsubsection{Time-Frequency Transformation -- STFT vs. fbsp-wavelets}
\label{sec:results:compare:tf_transform}

The evaluation results of the proposed fbsp-layer in comparison to the STFT demonstrate that it in general improves results (\autoref{tbl:results_ablation}).
Pre-training on AudioSet as an intermediate stage is desirable for the fbsp-based models if transfer learning is performed.
Absence of a such smooth transition between domains restricts the model performance on the target datasets.
Thus, the best performing setup includes our proposed fbsp-layer and an intermediate pre-training on AudioSet after the ImageNet initialization.

\subsubsection{Model -- ESResNe(X)t-fbsp vs. Others}
\label{sec:results:compare:others}

The proposed ESResNeXt-fbsp model achieves an outstanding accuracy on both datasets \mbox{ESC-50} (95.20\,\%) and UrbanSound8K (89.14\,\%).
In comparison to other approaches, it does not require the use of meta-learning \cite{kumar2020weanet} or ensembling \cite{palanisamy2020densenet} techniques in order to perform the best.
Moreover, the proposed model provides the highest single-model accuracy among the models that were fine-tuned on both target datasets (\autoref{tbl:results_comparison}).
Unlike many other models, our proposed fbsp-layer provides also insights on the desired by models representation of an input signal.

\begin{figure}[tbp]
	\begin{subfigure}[c]{\linewidth}
	    \centering
    	\includegraphics[trim={0cm 0.3cm 0cm 1cm},clip,width=\linewidth]{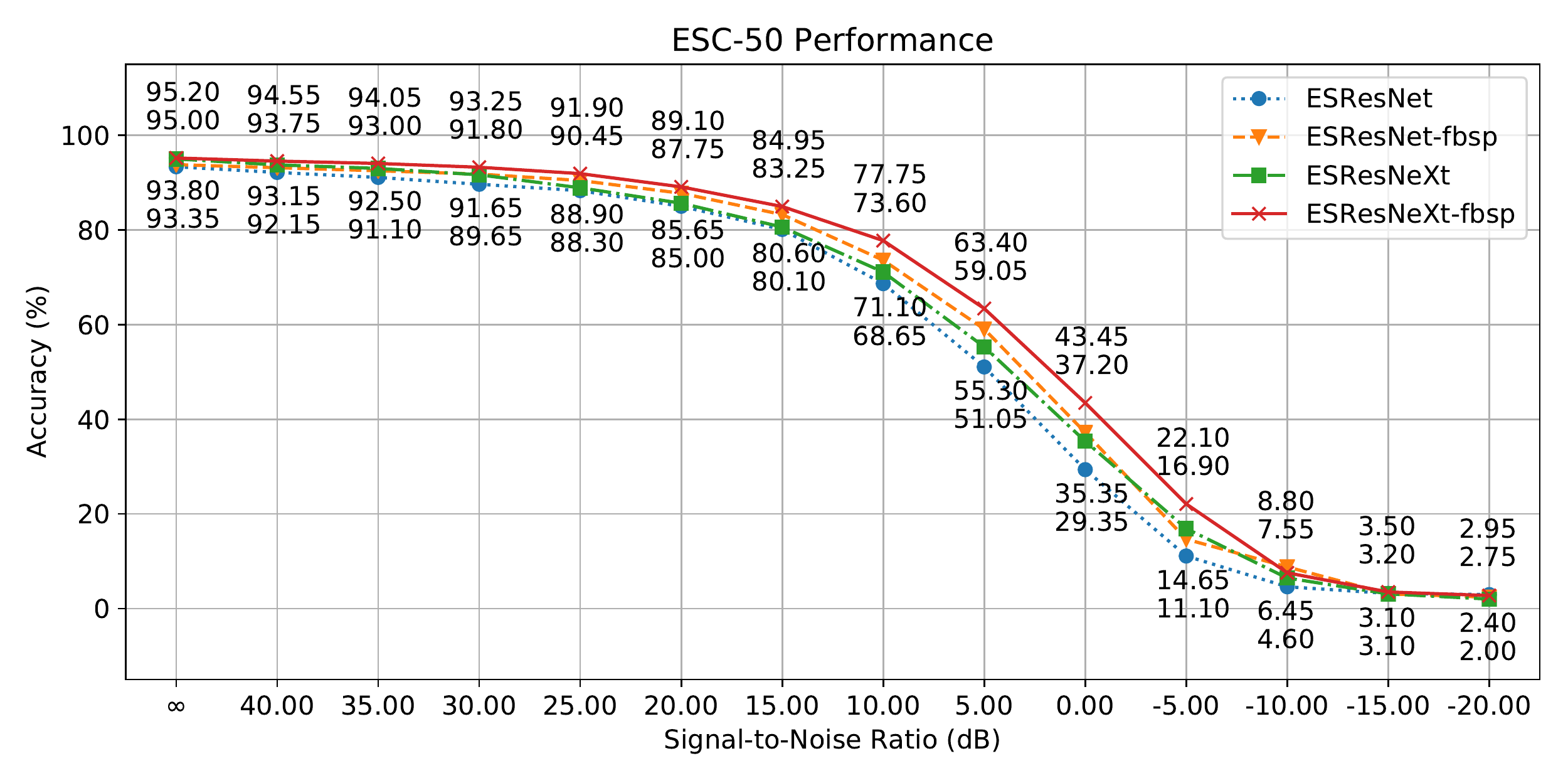}
    	\caption{\centering ESC-50}
    	\label{fig:robustness_awgn_esc50}
    \end{subfigure}
    
    \vspace{1em}
    \begin{subfigure}[c]{\linewidth}
	    \centering
    	\includegraphics[trim={0cm 0.3cm 0cm 1cm},clip,width=\linewidth]{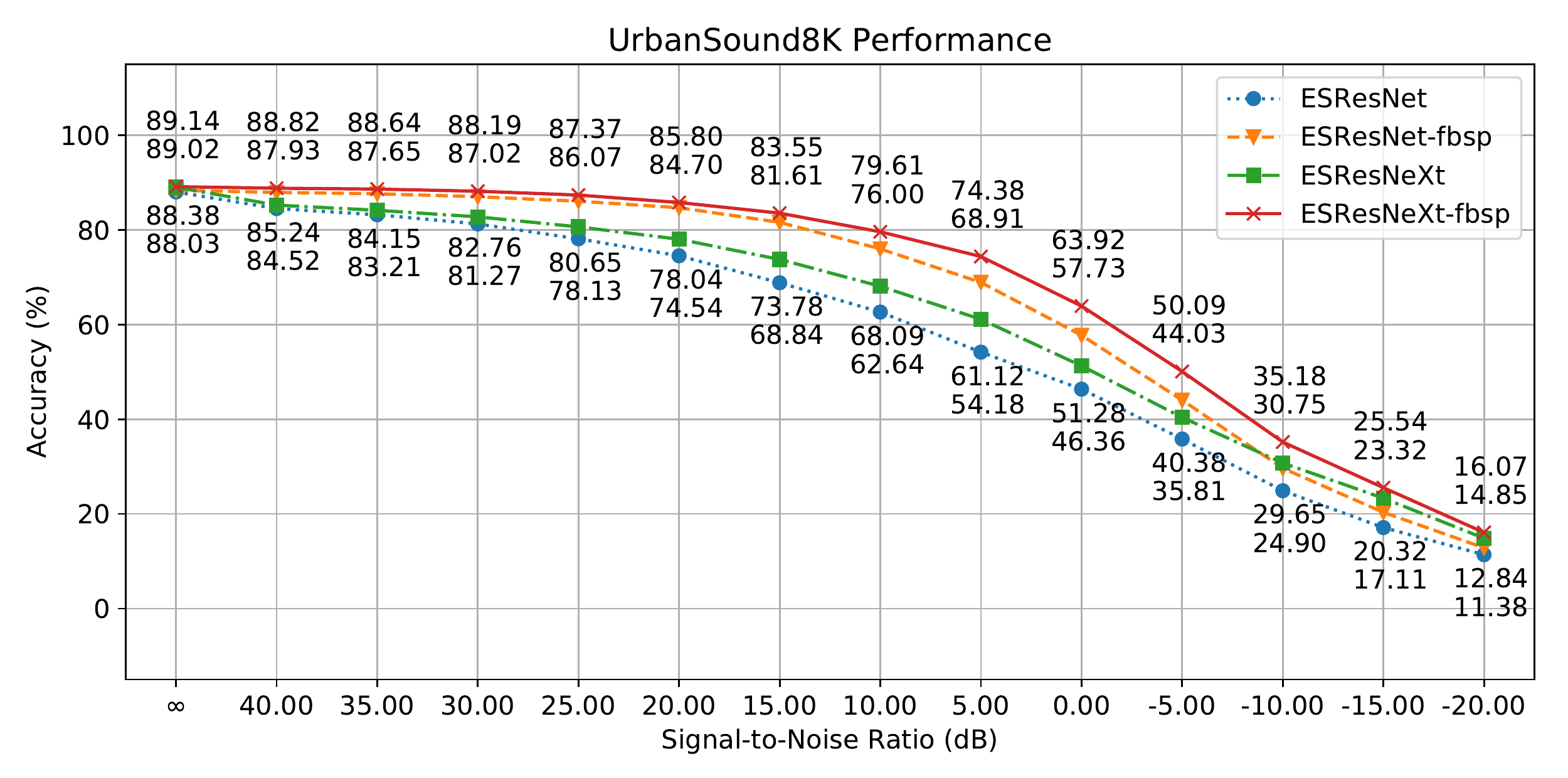}
    	\caption{\centering UrbanSound8K}
    	\label{fig:robustness_awgn_us8k}
    \end{subfigure}
    
	\caption{Dependency of model performance on the addition of white Gaussian noise. We can see that fbsp-based models are more robust against such a noise.}
	\label{fig:robustness_awgn}
\end{figure}

\begin{figure}[htbp]
	\begin{subfigure}[c]{\linewidth}
	    \centering
    	\includegraphics[trim={0cm 0cm 0cm 1cm},clip,width=\linewidth]{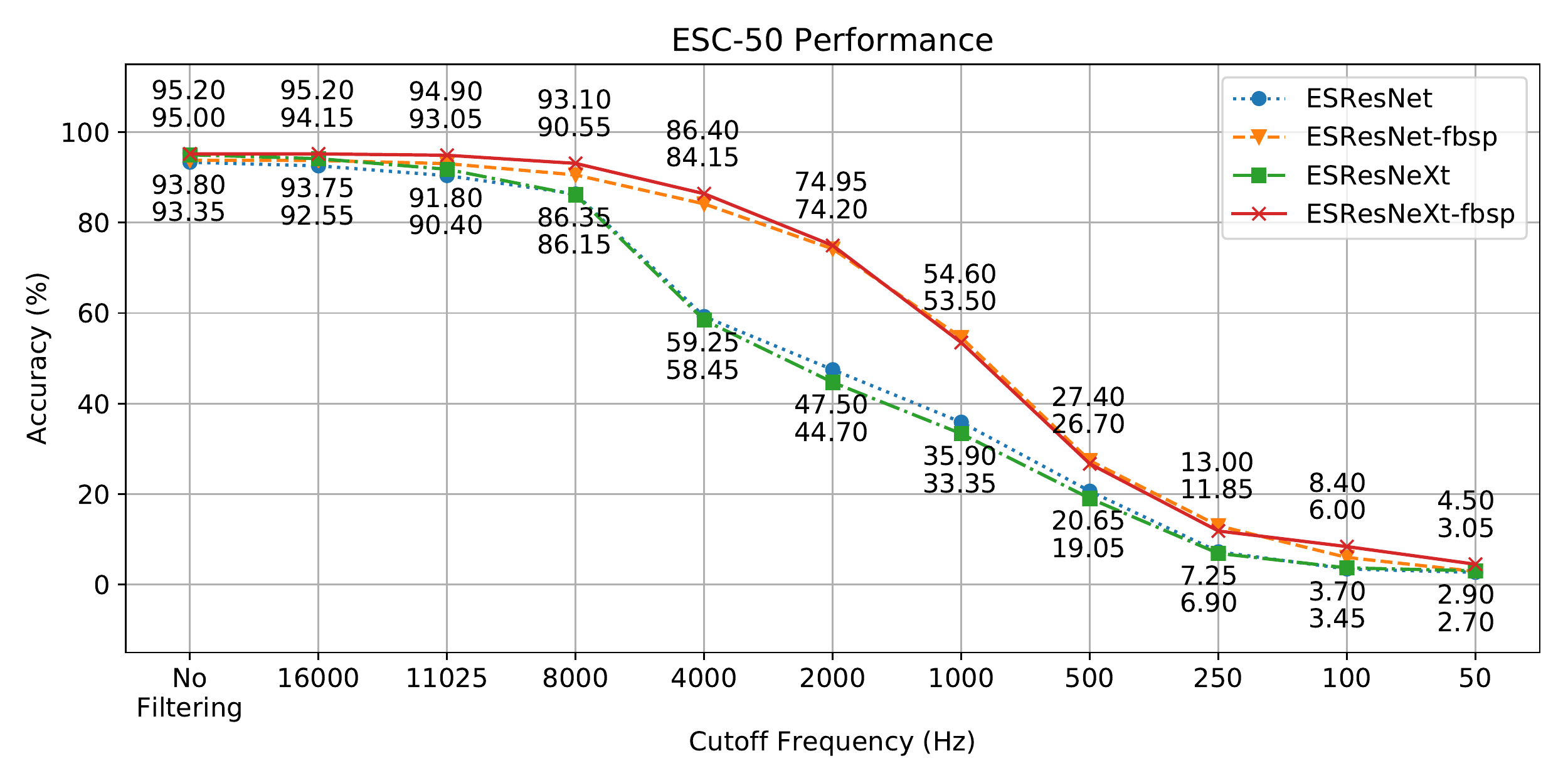}
    	\caption{\centering ESC-50}
    	\label{fig:robustness_sr_esc50}
    \end{subfigure}
    
    \vspace{1em}
    \begin{subfigure}[c]{\linewidth}
	    \centering
    	\includegraphics[trim={0cm 0cm 0cm 1cm},clip,width=\linewidth]{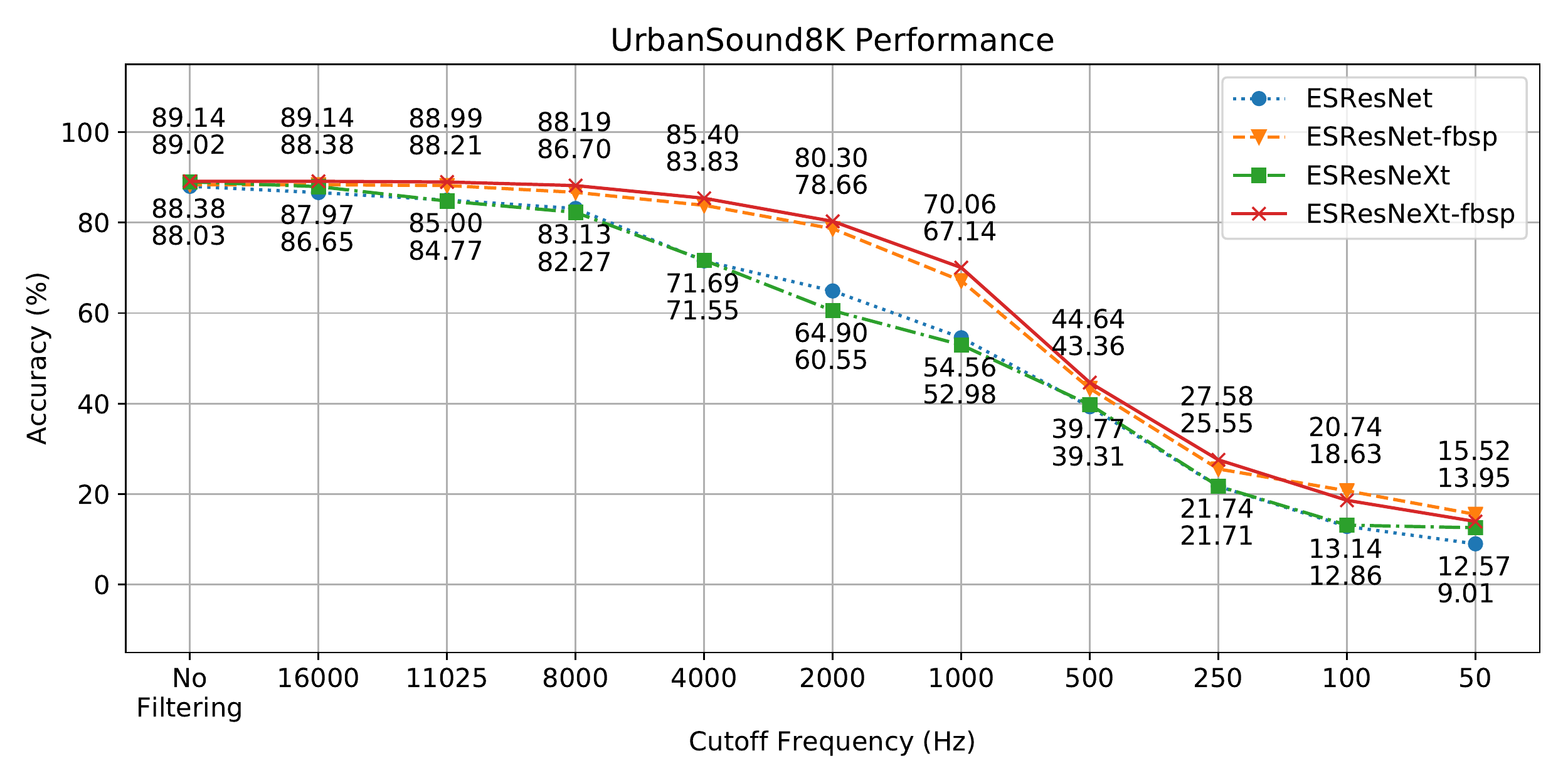}
    	\caption{\centering UrbanSound8K}
    	\label{fig:robustness_sr_us8k}
    \end{subfigure}
    
	\caption{Dependency of model performance on the effective sample rate of the input signal. We can see that fbsp-based models are more robust against lower frequency cutoffs.}
	\label{fig:robustness_sr}
\end{figure}

\begin{table*}[tbp]
\begin{threeparttable}[t]
\caption{Evaluation Results, accuracy (\%)}
\label{tbl:results_comparison}
\ra{1.2}
\begin{tabularx}{\textwidth}{@{}cXclcccc@{}}
\toprule
 & \multicolumn{1}{l}{\multirow{2}{*}{Model}} & \multicolumn{1}{c}{\multirow{2}{*}{Source}} & \multicolumn{1}{l}{\multirow{2}{*}{Representation}} & \multicolumn{1}{c}{ImageNet} & \multicolumn{1}{c}{AudioSet} & \multicolumn{1}{c}{\multirow{2}{*}{ESC-50}} & \multicolumn{1}{c}{\multirow{2}{*}{US8K}} \\
 & \multicolumn{3}{c}{} & \multicolumn{1}{c}{Initialized} & \multicolumn{1}{c}{Pre-Trained} & \multicolumn{2}{c}{} \\
\midrule
\parbox[t]{2mm}{\multirow{27}{*}{\rotatebox[origin=c]{90}{\underline{Others}}}} & Human (2015) & \cite{piczak2015esc} & -- & -- & -- & 81.30 & --\;\; \\
\cmidrule(l){2-8} 
 & \multicolumn{1}{l}{\textbf{Raw waveform and 1D-CNN}} & \multicolumn{6}{l}{} \\
 & EnvNet (2017) & \cite{tokozume2017envnet} & raw & & & 74.10 & 71.10\;\; \\
 & EnvNet v2 (2017)  & \cite{tokozume2017envnetv2} & raw & & & 84.70 & 78.30\;\; \\
 & Multiresolution 1D-CNN (2018)  & \cite{zhu2018multires} & raw & & & 75.10 & --\;\; \\
\cmidrule(l){2-8} 
 & \multicolumn{1}{l}{\textbf{Time-frequency representation and 2D-CNN}} & \multicolumn{6}{l}{} \\
 & Piczak-CNN (2015)  & \cite{piczak2015cnn} & Mel-spec & & & 64.50 & 73.70\;\; \\
 & SB-CNN (2017)  & \cite{salamon2017cnn} & Mel-spec & & & -- & 79.00\;\; \\
 & Piczak-CNN (2017)  & \cite{tak2017pefbe} & (PE)FBE & & & 84.15 & --\;\; \\
 & VGG-like CNN + mix-up (2018)  & \cite{zhang2018mixup} & Mel-, GT-spec & & & 83.90 & 83.70\;\; \\
 & VGG-like CNN + Bi-GRU + attention (2019)  & \cite{zhang2019crnn} & GT-spec & & & 86.50 & --\;\; \\
 & CNN10 (2020)  & \cite{arnault2020urban} & Mel-spec & & \checkmark & 90.00 & 86.10\;\; \\
 & WEANET $N^{0}$ (2020)  & \cite{kumar2020weanet} & Mel-spec & & \checkmark & 92.60 & --\;\; \\
 & WEANET $N^{4}$ (2020)  & \cite{kumar2020weanet} & Mel-spec & & \checkmark & 94.10 & --\;\; \\
 & DenseNet-201 (2020)  & \cite{palanisamy2020densenet} & Mel-spec & & & 72.50 & 76.32\;\; \\
 & DenseNet-201 (2020)  & \cite{palanisamy2020densenet} & Mel-spec & \checkmark & & 91.16 & 85.14\;\; \\
 & DenseNet-201$\,\times\,5$, ensemble (2020)  & \cite{palanisamy2020densenet} & Mel-spec & \checkmark & & 92.89 & 87.42\;\; \\
 \cmidrule(l){2-8}
 & \multicolumn{1}{l}{\textbf{Trainable filter bank and 2D-CNN}} & \multicolumn{6}{l}{} \\
 & Piczak-CNN + ConvRBM (2017) & \cite{sailor2017convrbm} & FBE & & & 86.50 & --\;\; \\

\cmidrule(l){1-8} 
\parbox[t]{2mm}{\multirow{6}{*}{\rotatebox[origin=c]{90}{\underline{Ours}}}} & \multicolumn{7}{l}{\textbf{Time-frequency representation and 2D-CNN}} \\
 & ESResNet & & STFT-spec & \checkmark & \checkmark & 93.35 & 88.03\;\; \\
 & ESResNeXt & & STFT-spec & \checkmark & \checkmark & 95.00 & 89.02\;\; \\
 & \multicolumn{7}{l}{\textbf{Learnable filterbank and 2D-CNN}} \\
 & ESResNet-fbsp & & fbsp-spec & \checkmark & \checkmark & 93.80 & 88.38\;\; \\
 & ESResNeXt-fbsp & & fbsp-spec & \checkmark & \checkmark & \textbf{95.20} & \textbf{89.14}\;\; \\
\bottomrule
\end{tabularx}
\begin{tablenotes}
\footnotesize
    \underline{Abbreviations}:\:
    \item FBE: FilterBank Energies \cite{sailor2017convrbm};
    \item spec: spectrogram;
    \item GT: GammaTone \cite{slaney1993gammatone};
    \item (PE)FBE: (Phase-Encoded) FBE \cite{tak2017pefbe};
    \item STFT: Short-Time Fourier Transform;
    \item fbsp: (complex) Frequency B-SPline (wavelets).
\end{tablenotes}
\end{threeparttable}
\end{table*}

\subsection{Model Robustness}
\label{sec:results:robustness}
Apart from the model comparison, we also evaluated the robustness against two types of signal perturbations: additive white Gaussian noise and reduction of an effective sample rate.

\subsubsection{Additive White Gaussian Noise}
\label{sec:results:robustness:awgn}
The evaluation of model performance on the ESC-50 and UrbanSound8K datasets given different values of SNR shows clearly (\autoref{fig:robustness_awgn}) that the use of the fbsp-layer improves the model's robustness to the presence of additive white Gaussian noise.
As indicated in \autoref{fig:freq_resp_esrn_fbsp_as_fs}\:--\:\autoref{fig:freq_resp_esrnx_fbsp_as_ptin}, the higher frequency band is being suppressed by trained fbsp-layers, in comparison to the DFT-filter bank.
This allows to reduce the amount of the added noise partially, thus, improving the signal-to-noise ratio for the obtained spectrograms.
As shown in \autoref{fig:robustness_awgn}, fbsp-equipped models provide higher classification accuracy given decreasing SNR on both datasets, \mbox{ESC-50} and UrbanSound8K.

\subsubsection{Reduction of an Effective Sample Rate}
\label{sec:results:robustness:sr}

In order to quantify the influence of the effective sample rate reduction on the model accuracy, several experiments were performed.
The obtained results support the point that fbsp-equipped models are able to extract the information that is relevant for classification more effectively.
In particular, the significant performance drop at the cutoff frequency $4\:\si{\kilo\hertz}$ occurred in the case of the STFT-based models on both datasets (\autoref{fig:robustness_sr}) indicates that the use of the fbsp-equipped models is beneficial for the applications that imply the use of low-bandwidth channels. Also, the ResNet backbone demonstrated less sensitivity to lower frequency cutoffs, in comparison to ResNeXt (\autoref{fig:robustness_sr}).

\section{Conclusion}
\label{sec:conclusion}
In this paper, we proposed a new fbsp-layer that is based on complex frequency B-spline wavelets and tailored towards effective and robust time-frequency representation.

Based on the fbsp-layer, and a common ResNeXt, our ESResNeXt-fbsp model achieves new state-of-the-art results on two datasets: ESC-50 (95.20\,\%) and UrbanSound8K (89.14\,\%).
To ease reproducibility, detailed code and settings of our approach are published\footnote{\url{https://github.com/AndreyGuzhov/ESResNeXt-fbsp}}.

We also evaluated the influence of the additional pre-training on the AudioSet dataset, which is beneficial for the model performance, as well as the improvement obtained by the change of the model's backbone from ResNet-50 to ResNeXt-50.

Further, we found that the proposed fbsp-layer allows to obtain ESC models that are more robust against additive white Gaussian noise and a possible reduction of the sample rate, in comparison to models that were trained using STFT-based spectrograms as an input.
Additionally, the frequency responses of the trained fbsp-filter banks provide insights into the importance of specific frequencies for the audio classification, making it possible to understand the models' predictions and behavior better.

In the future, we would like to further investigate the influence of the internal cardinality of the fbsp-layer, as an increased number of internal parameters could potentially further improve model performance.
Also, changing the current split of input spectrograms according to RGB-channels to a full-frame representation could influence positively on the prediction accuracy, so we would like to quantify its effect.

\section*{Acknowledgments}
We thank all members of the Deep Learning Competence Center at the DFKI for their comments and support.

\bibliographystyle{IEEEtran}
\bibliography{references}

\end{document}